\documentclass{emulateapj}  
\usepackage{epsf}
\usepackage{subfigure}
\usepackage{latexsym}  
\usepackage{epsfig}
\usepackage{natbib}

\def\msun{{\rm M}_\odot}
\def\beq{\begin{equation}}
\def\eeq{\end{equation}}

\def\rvir{R_{\rm vir}}

\def\mvir{M_{\rm vir}}

\def\mg500{M_{g,500}}
\def\epsdm{\epsilon_{\rm DM}}
\def\epsfb{\epsilon_{\rm f}}
\def\mhse{M_{500}^{\rm HSE}}
\def\m500{M_{500}}
\def\r500{R_{500}}

\def\fgas{f_g}
\def\hmsun{\; h^{-1} M_{\odot}}
\def\h70msun{\; h_{70}^{-1} \; M_{\odot}}

\def\dell{\mathcal{D}_\ell}

\begin{document}

\bibstyle{aas}

\shorttitle{Impact of Cluster Physics on the SZ Power Spectrum}
\shortauthors{Shaw et al.}

\author{Laurie D. Shaw\altaffilmark{1,2}, Daisuke Nagai\altaffilmark{1,2},  Suman Bhattacharya\altaffilmark{3}, Erwin T. Lau\altaffilmark{4}}
\altaffiltext{1}{Department of Physics, Yale University, New Haven, CT 06520}
\altaffiltext{2}{Yale Center for Astronomy \& Astrophysics, Yale University, New Haven, CT 06520}
\altaffiltext{3}{T-2, Theoretical Division, Los Alamos National Laboratory, Los Alamos, NM 87545}
\altaffiltext{4}{Department of Astronomy \& Astrophysics, 5640 South Ellis Ave., The University of Chicago, Chicago, IL 60637 }
\email{laurie.shaw@yale.edu}

\title{Impact of Cluster Physics on the Sunyaev-Zel'dovich Power Spectrum}

\begin{abstract}
We use an analytic model to investigate the theoretical uncertainty on
the thermal Sunyaev-Zel'dovich (SZ) power spectrum due to
astrophysical uncertainties in the thermal structure of the
intracluster medium. Our model accounts for star formation and energy
feedback (from supernovae and active galactic nuclei) as well as
radially dependent non-thermal pressure support due to random gas
motions, the latter calibrated by recent hydrodynamical
simulations. We compare the model against X-ray observations of low
redshift clusters, finding excellent agreement with observed pressure
profiles. Varying the levels of feedback and non-thermal pressure
support can significantly change both the amplitude and shape of the
thermal SZ power spectrum. Increasing the feedback suppresses power at
small angular scales, shifting the peak of the power spectrum to lower
$\ell$. On the other hand, increasing the non-thermal pressure support
has the opposite effect, significantly reducing power at large angular
scales. In general, including non-thermal pressure at the level
measured in simulations has a large effect on the power spectrum,
reducing the amplitude by 50\% at angular scales of a few arcminutes
compared to a model without a non-thermal component. Our results
demonstrate that measurements of the shape of the power spectrum can
reveal useful information on important physical processes in groups
and clusters, especially at high-redshift where there exists little
observational data.  Comparing with the recent South Pole Telescope
measurements of the small-scale cosmic microwave background power
spectrum, we find our model reduces the tension between the values of
$\sigma_8$ measured from the SZ power spectrum and from cluster
abundances.
\end{abstract}
\keywords{cosmology: dark matter --- galaxies: clusters: general --- 
intergalactic medium}

\section{INTRODUCTION}

The Sunyaev-Zel'dovich (SZ) effect has long been recognized as a
powerful tool for probing the physics of the intracluster medium
(ICM), large-scale structure formation and the dark energy equation of
state \citep{birkinshaw99, carlstrom02}. Experiments such as the South
Pole Telescope \citep{ruhl04}, the Atacama Cosmology Telescope
\citep{kosowsky03} and
Planck\footnote{http://www.rssd.esa.int/index.php?project=Planck} are
currently surveying the microwave sky with the goal of
identifying clusters via their SZ signature on the cosmic microwave
background (CMB), thus constructing large catalogs of galaxy clusters
that are uniformly selected by SZ flux \citep{vanderlinde10}.

The SZ effect can also be detected as a secondary anisotropy in the
CMB temperature power spectrum, appearing as ``excess power'' (over the
predicted primary anisotropy signal) on angular scales of a few
arc-minutes. The ensemble averaged power spectrum amplitude $\bar
C_\ell$ has an extremely sensitive dependence on $\sigma_8$, $\bar
C_\ell \propto \sigma_8^7(\Omega_b h)^2$ \citep{komatsu02}, where
$\sigma_8$ is the power spectrum normalization on scales of
$8h^{-1}$~Mpc. The SZ angular power spectrum thus represents a robust
observable with which competitive constraints on $\sigma_8$ can be
obtained. This would in turn enable tighter constraints to be placed
on the dark energy equation of state parameter $w$ by breaking the
degeneracy with $\sigma_8$ in constraints derived from the primary CMB
signal alone.

A number of experiments have reported detections of power in excess of
the primary CMB power spectrum at small angular scales and thus been
able to place upper limits on the amplitude of the SZ signal
\citep{dawson06, friedman09, reichardt09, reichardt09b, sievers09,
  sharp10, fowler10}.  The South Pole Telescope
\citep[SPT,][]{lueker10} has recently reported measurements of the CMB
power spectrum at 150 GHz for angular scales in the range $2000 < \ell
< 10000$. By combining SPT survey maps at 150 and 220 GHz to minimize
astrophysical foreground signals, \citet{lueker10} were able to
isolate and detect SZ power (kinetic plus thermal) at $2.6\sigma$. The
measured amplitude at $\ell=3000$ was $4.2 \pm 1.5 \mu K^2$,
significantly below that predicted by halo model calculations
\citep{komatsu02} or simulations \citep{white02, shaw09, sehgal10},
assuming WMAP7 cosmological parameters. The significantly
lower-than-predicted signal could be explained by a lower value of
$\sigma_8$.  Using the SZ power spectrum predicted by simulations and
assuming no modeling uncertainty, \citet{lueker10} measured $\sigma_8
= 0.746 \pm 0.017$. However, this result is in tension with other
probes of $\sigma_8$; for example, recent X-ray and optical
measurements of cluster abundances suggest $\sigma_8 = 0.82 \pm 0.05$
\citep{mantz10}, $\sigma_8(\Omega_M/0.25)^{0.47} = 0.813 \pm 0.013$
\citep{vikhlinin09}, and $\sigma_8 = 0.807 \pm 0.02$ \citep{rozo10}.

This disagreement can be resolved if current analytic models and
simulations over-predict the level of SZ power by a factor of $\approx
2$. SZ power spectrum calculations have two sources of uncertainties:
the amplitude of the halo mass function, and in modeling the radial
electron pressure profile of the ICM. Recent studies of the mass and
redshift distribution of halos in N-body simulations imply that the
mass function is known to 5\%-10\% accuracy for the currently allowed
wCDM cosmology \citep{tinker08, bhattacharya10} and so is not the
dominant theoretical uncertainty \citep[although, as pointed out by][
  the impact of baryonic physics on halo masses is still somewhat
  unclear]{stanek09}.

The main difficulty in calculating the thermal SZ (tSZ) power spectrum
is predicting the thermal pressure profiles of groups and clusters
over a wide range of mass and redshift. \citet{komatsu02} and
\citet{holder02} show that low-mass ($M<2\times 10^{14} \hmsun$) and
high-redshift ($z > 0.6$) objects both make a significant contribution
to the signal at angular scales of a few arcminutes ($\ell \approx
4000$), the scales at which current small-scale anisotropy experiments
such as SPT are most sensitive. While {\it Chandra} and {\it
  XMM-Newton} have enabled high-mass, low-redshift clusters to be
studied extensively over the last decade, lower mass and, in
particular, high-redshift objects have not been extensively
studied. Furthermore, the SZ power spectrum is sensitive to the
thermal pressure of the ICM out to cluster radii several times larger
than is typically probed by X-ray observations.  Hence, there are few
observational constraints that can be used to aid predictions of the
tSZ power spectrum.

From a theoretical perspective, full cosmological hydrodynamical
simulations have only recently begun to systematically explore the
effects of sub-grid baryonic processes, such as radiative cooling,
star formation, feedback mechanisms, cosmic rays, thermal conduction
and magnetic fields, on the thermal structure of the ICM
\citep[][]{nagai07,pfrommer07,dolag09,sijacki07, sijacki08,
  battaglia10}.  Furthermore, simulations require high spatial
resolution in order to effectively implement some of these
processes. However, large box sizes are also required in order to
adequately sample the halo mass function at group and cluster scales
to enable measurements of the tSZ power spectrum. Currently, the
computational expense of running large box, high-resolution
hydrodynamical simulations is prohibitive to investigating the level
of theoretical uncertainty on the power spectrum as well as the
dependence on cosmological parameters.

The principal aim of this work is to use analytic models to
investigate variations in the predicted tSZ power spectrum caused by
uncertainties in the thermal structure of the ICM. Specifically, we
study the impact of energy feedback, non-thermal pressure support and
halo concentration on the shape and amplitude of the power
spectrum. This is achieved by combining an analytic model for the ICM
with the halo mass function to rapidly calculate the power spectrum
for different model parameters. Our model assumes that gas resides in
hydrostatic equilibrium with a polytropic equation of state, and
accounts for star-formation as well as feedback from supernovae and
active galactic nuclei (AGNs) and non-thermal pressure support driven
by random gas motions and turbulence in the ICM. We calibrate our
model parameters by comparing against X-ray observations of massive,
low-z clusters. We find that including non-thermal pressure support at
the level measured in state-of-the-art hydrodynamical simulations
significantly reduces the amplitude of the predicted tSZ power
spectrum, thus reducing the tension between the $\sigma_8$ inferred
from the SPT observations and cluster abundance measurements.

The paper is organized as follows. In Section~\ref{sec:model}, we
describe our model of the ICM. In Section~\ref{sec:calibration}, we
explore the model parameter space by comparing model scaling relations
and radial profiles against recent low-redshift X-ray observations of
groups and clusters.  In Section~\ref{sec:szpowerspec}, we explore the
uncertainty on the tSZ power spectrum sourced by the underlying range
in ICM model parameters. We compare our fiducial power spectrum model
with other recent simulations, and discuss our results in the context
of the recent SPT observations.

Throughout this paper we assume a fiducial, spatially flat,
$\Lambda$CDM cosmological model consistent with the WMAP7 best-fit
cosmological parameters, namely $H_0 = 71$ km s$^{-1}$Mpc$^{-1}$,
$\Omega_M = 0.264$, $\Omega_b = 0.044$, $\Omega_\Lambda = 0.736$, $n_s
= 0.96$ and $\sigma_8 = 0.8$.

\section{Theoretical Models}
\label{sec:model}

\subsection{Thermal SZ Power Spectrum}
\label{sec:tszmodel}

The tSZ effect is a distortion of the CMB caused by inverse
Compton scattering of CMB photons off electrons in the high temperature
plasma within galaxy clusters. To first order, the temperature change
at frequency $\nu$ of the CMB is given by $\Delta T/T_{\rm CMB} (x_{\nu})
= f(x_\nu) y$, where $f(x_\nu) = x_\nu(\coth(x_\nu/2) - 4)$, $x_\nu = h\nu /
k_B T_{\rm CMB}$, and $y$ is the dimensionless Compton-$y$ parameter
\begin{equation}
y = \left(\frac{k_B \sigma_T}{m_e c^2}\right)\int n_e(l)T_e(l) dl \;,
\end{equation}
where the integral is along the line of sight, $T_{\rm CMB}$ is
the CMB temperature, and $n_e$ and $T_e$ are the number density
and electron temperature of the ICM, respectively. 

The tSZ power spectrum can be calculated by simply summing up the
squared Fourier-space SZ profiles of all clusters:
\begin{equation}
C_{\ell} =  f(x_\nu)^2\int dz {dV \over dz } \int d \ln M {dn(M,z) \over d \ln M}
\tilde{y}^2(M,z,\ell)
\label{eq:tsz_powerspec}
\end{equation}
where $V(z)$ is the comoving volume per steradian and $n(M,z)$ is the
number density of objects of mass $M$ at redshift $z$. For the latter
we use the fitting function of \citet{tinker08}. $\tilde{y}(M,z,\ell)$
is the Fourier transform of the projected SZ profile for a cluster of
mass $M$ and redshift $z$. This can be calculated assuming spherical
symmetry using \citep{bracewell00, komatsu02}:
\beq
\tilde{y}(M,z,\ell) = \frac{4 \pi r_c}{\ell_c^2}\int_0^\infty dx x^2 P_y(M,z,x) \frac{\sin(\ell x/\ell_c)}{\ell x / \ell_c} \;\;, 
\eeq
where $x = r/r_c$, $\ell_c = D_A(z)/r_c$, $r_c$ is a characteristic scale
radius of the profile, $D_A(z)$ is the angular diameter distance to redshift $z$ and
$p_y(M,z,x)$ is the three-dimensional SZ profile, which is related simply to the gas thermal electron 
pressure profile,
\beq
P_y(M,z,x) \equiv \frac{\sigma_{\rm T}}{m_e c^2}  P_e(M,z,x) \;.
\eeq
Note that while this calculation only accounts for the one-halo
contribution to $C_{\ell}$, \citet{komatsu99} demonstrated that, at
the angular scales being probed by the current generation of small-scale CMB
experiments (i.e., $\ell \geq 1000$), the two-halo (or clustered)
contribution to the tSZ power spectrum is nearly 2 orders of
magnitude smaller than the one-halo term and decreases with increasing
$\ell$. We thus neglect the two-halo contribution in our calculation.

Groups and clusters over a wide mass ($10^{13} < {\rm M}/\msun <
10^{15}$) and redshift ($0<z<3$) range contribute to the tSZ power
spectrum. For example, more than half the power at $\ell = 3000$ comes
from clusters at redshift greater than one \citep{komatsu02}.
Likewise, at the same angular scale, half of the power is predicted to
be sourced by objects of mass less than $2 \times 10^{14}
\msun$. While the mass and redshift contribution depends on the
details of the underlying gas physics incorporated in models and
simulations, it is clear that much of the signal comes from objects
for which there exists little direct observational data.

\subsection{Cluster Model}
\label{sec:icm_mod}

Our model for the density and temperature structure of the ICM is
based upon the model of \citet[][henceforth O05]{ostriker05} and
\citet[][henceforth B09]{bode09}, with an important modification that
allows for a radially-dependent non-thermal pressure component
(e.g., turbulence or bulk flows). In this section, we briefly review the
model and describe how non-thermal pressure support is implemented; we
refer the reader to O05 and B09 for a more detailed description of the
original model.

\subsubsection{Dark Matter Halo Structure}
\label{sec:dmhalo}

The model assumes that the ICM initially follows the density
and temperature of the host dark matter halo, but rapidly rearranges
into hydrostatic equilibrium within the potential well of the host
dark matter halo with a polytropic equation of state. 

Dark matter halo properties are determined by a Navarro-Frenk-White (NFW) density 
profile \citep{navarro96,navarro97} of the form,
\beq
\label{eq:nfw}
\rho_{DM}(r) =  \frac{\rho_s}{x (1+x)^{2}} \;\;,
\eeq
where $x\equiv r/r_{s}$, $r_s$ is the NFW scale radius, and $\rho_s$ is
a normalization constant. Halo concentration is defined as $c \equiv \rvir/r_s$. 
Numerous N-body simulations have demonstrated there 
to be a power-law scaling between halo mass, redshift and concentration. 
We adopt the halo mass-concentration relation measured by \citet{duffy08} 
from halos identified in a large N-body simulation over the
redshift range $0 \leq z \leq 2$,
\beq 
c(M,z) = 7.85  A_C \left(\frac{\mvir}{2\times10^{12} \hmsun}\right)^{-0.081} (1+z)^{-0.71}
\;,  
\label{eq:massconc}
\eeq where $A_C$ is an arbitrary normalization factor that we will
later vary to investigate the impact of the mass-concentration
normalization on the tSZ power spectrum through our model.  
$A_C = 1$ gives the mass-concentration relation of \citet{duffy08}.  

Halo masses are defined in terms of the virial overdensity;
\beq 
\mvir = \frac{4}{3} \rvir^3 \Delta_c \rho_c(z) \;,
\eeq
where $\Delta_c = 18 \pi^2 + 82 (\Omega_M(z)-1) -39 (\Omega_M(z)-1)^2$
is the virial overdensity given in \cite{bryan98} and $\rho_c(z)$ is
the critical density at redshift $z$. When comparing with
observations, we also use the mass definition $\m500 = (4/3) \pi
\r500^3 500 \rho_c$, where $\r500$ is the radius out to which 
current X-ray observations can reliably measure the gas density
and temperature profiles. 

\subsubsection{Star Formation}
\label{sec:baryons}

The gas is initially treated as a tracer of negligible mass of the dark
matter gravitational potential, i.e., it does not contribute itself to
the gravitational potential of the cluster and the gas density and
temperature follows that of the dark matter. 

We assume that some fraction of the gas has radiatively cooled and
formed stars. To determine the stellar mass of a cluster we use the
observed scaling relation of \citet{giodini09}, who measured the ratio
of stellar mass to total mass for 91 X-ray selected groups and
low-mass clusters in the COSMOS survey \citep{scoville07}. This was
supplemented by the 27 low-redshift, X-ray selected clusters analyzed
by \citep{lin03}, resulting in an overall sample that encompassed a
wide range of both mass ($10^{13} \leq \m500 \leq 10^{15}$) and
redshift ($0 \leq z \leq 1$). They found that the stellar mass
fraction within $\r500$, $f_*(<\r500) = M_*(<\r500)/M_{500}$, followed
a mean empirical relation; 
\beq 
f_* = (2.58\pm0.05)\times 10^{-2}
\left( \frac{M_{500}} {3\times 10^{14} \;\msun} \right)^{-0.37\pm0.04}\;\;.
\label{eq:stellarmass}
\eeq 

We assume that the stellar mass fraction given above also holds within
$\rvir$, i.e., $f_* = M_*(<\r500)/\m500 = M_*(<\rvir)/\mvir$. Following
O05 and B09, we determine the redshift evolution of the stellar mass
fraction adopting the ``fossil'' model of \citet{nagamine06}. In this
model the star-formation rate is given by a delayed exponential, with
a decay time of $1.5$ Gyr for bulge populations and $4.5$ Gyr for disk
populations. This model predicts only a small ($\approx 15\%$)
decrease between $z = 0$ and $1$ in the normalization of this
relation, which is consistent with the results of \citet{giodini09}.

The initial total gas mass within the virial radius is thus 
\beq
M_{g,i} = (f_b - f_*)\mvir
\label{eq:initmass}
\eeq
and the initial total energy of the gas is
\beq
E_{g,i} = f_b \left[2 \pi \int_{r_*}^{\rvir} \rho_{DM}(r)3\sigma^2_{DM}(r)r^2 dr + \int_{r_*}^{\rvir} \Phi(r) \frac{dM}{dr}dr \right]\;,
\label{eq:initenergy}
\eeq where $f_b = \Omega_b / \Omega_M$, $\Phi(r)$ is the gravitational
potential and $\sigma_{DM}(r)$ is the one-dimensional dark matter
velocity dispersion \citep[see, for example, Equation (13) of][and we
  assume isotropic dark matter orbits]{lokas01}. Thus, the initial
total energy of the gas is simply the sum of the kinetic and potential
energy of the dark matter halo scaled by the cosmic baryon
fraction. We assume that the gas within a radius $r_*$ has cooled and
formed stars, where $r_*$ is determined by $f_b M_{DM}(<r_*) = f_*
\mvir$ and so omit the contribution to the total gas energy within
this radius ($r_* \approx 0.24$ Mpc for a $2 \times 10^{14} \hmsun$
cluster at $z = 0.5$, decreasing weakly toward higher redshift/lower
mass). We note that the final gas mass within $\rvir$ predicted by our
model is not necessarily equal to $M_{g,i}$, as the gas may expand or
contract when hydrostatic equilibrium is enforced (see Section
\ref{sec:feedback}).

\subsubsection{Hydrostatic Equilibrium}
\label{sec:hse}

The ICM should rapidly rearrange itself into hydrostatic
equilibrium, satisfying
\beq
\frac{dP_{tot}(r)}{dr} = -\rho_g(r) \frac{d\Phi(r)}{dr} \;,
\eeq
where $\rho_g(r)$ is the gas density at radius $r$ from the cluster center and $P_{tot}$ is the
total gas pressure, $P_{tot}(r) = P_{th}(r) + P_{nt}(r)$. The total
gas pressure is therefore a combination of the thermal and non-thermal
pressure components, with the latter primarily due to random gas
motions and turbulence in the ICM. We discuss how we implement a
non-thermal pressure component in the following section.

\begin{figure}
\includegraphics[scale = 0.44]{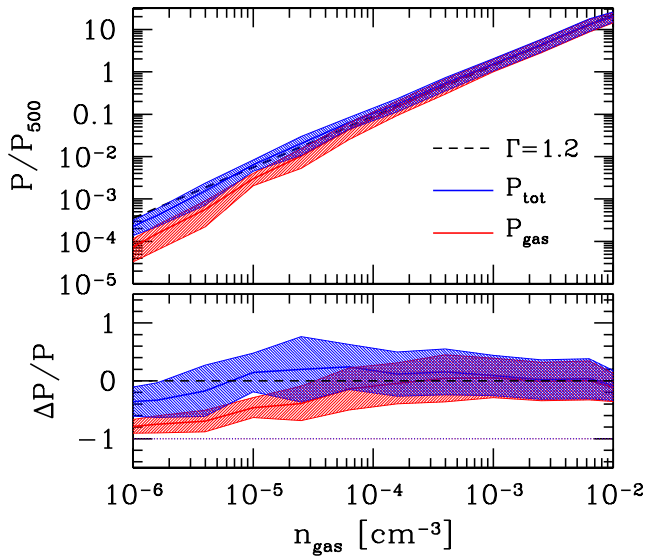}
\caption{Total (blue line) and thermal (red) pressure as a function of
  gas number density, averaged over 16 clusters simulated by
  \citet{nagai07} at $z=0$. The shaded regions show the standard
  deviation of the cluster sample around the mean. The solid black line
  shows the best fit line to the total pressure and density with a
  slope (adiabatic index) of $\Gamma = 1.2$. The lower panel shows the
  fractional deviation of the pressure around this line.}
\label{fig:phasediagram}
\end{figure}

Hydrodynamical simulations indicate that the {\it total} pressure
adheres more closely to a polytropic equation of state than the
thermal pressure, i.e., $P_{tot} \propto \rho_g^\Gamma$.  Figure
\ref{fig:phasediagram} shows the pressure-density phase diagram
obtained for 16 clusters simulated using the Eulerian hydrodynamics
ART code \citep{kravtsov02}, including radiative cooling and
star formation \citep{nagai07}. The total pressure (blue line) is the
sum of the thermal (red line) and non-thermal components in radial
bins around the cluster potential minimum, averaged over the entire
sample. The shaded regions denote the standard deviation within the
sample around the mean. The non-thermal pressure is measured from
the radial velocity dispersion of the gas within each shell
\citep{lau09}. The black dashed line represents $\Gamma = 1.2$.

It is clear from this figure that using the total pressure results in
a constant adiabatic index of 1.2 over more than four decades of gas
density, whereas the thermal pressure implies a density (and hence
radially) dependent $\Gamma$. The same results are obtained for
non-radiative simulations without gas cooling and star-formation.  We
therefore assume that the {\it total} gas pressure has a polytropic
equation of state, $P_{tot} = P_0 (\rho/\rho_0)^\Gamma$, where we set
$\Gamma = 1+1/n =1.2$ ($n =5$, where $n$ is the polytropic index) and
$\rho_0$ and $P_0$ are the central density and pressure of the gas. In
Section \ref{sec:nonthermpress} we describe our method for including
non-thermal pressure in the model and demonstrate that this causes the
ratio $P_{th} / \rho^\Gamma$ to vary with cluster radius.

The final total pressure and density of the gas in our model are given by
\begin{eqnarray}
P_{tot}(r) &=& P_o \theta(r)^{n+1}  \\
\rho_{g}(r) &=& \rho_0 \theta(r)^{n} \;,
\label{eq:model_press_den}
\end{eqnarray}
where $\theta(r)$ is the polytropic variable
\beq
\theta(r) = 1 + \frac{\Gamma-1}{\Gamma}\frac{\rho_0}{P_0}(\Phi_0 - \Phi(r)) \;,
\eeq
and $\Phi_0$ is the central potential of the cluster. 

\subsubsection{Energy Feedback}
\label{sec:feedback}

Once $P_0$ and $\rho_0$ are determined, the density and pressure
profile of the gas is fully specified. These variables are determined
by applying two constraints. First, the total final energy of the
gas that is initially within $\rvir$ must obey
\beq
E_{g,f} = E_{g,i} + \epsdm |E_{DM}| + \epsfb M_* c^2 + \Delta E_p \;.
\eeq
This condition states that the final energy $E_{g,f}$ be equal to the
initial energy, $E_{g,i}$ (Equation \ref{eq:initenergy}), plus energy
added to the gas via dynamical process or from energy feedback
(discussed below), plus the work done by the gas as it expands or
contracts relative to its initial state. We define $R_f$ as the radius
within which the final gas mass, following polytropic rearrangement,
is equal to $M_{g,i}$ (Equation \ref{eq:initmass}).  The work done by
the gas is then given by $\Delta E_p = (4\pi/3)(\rvir^3 -
R_{f}^3)P_{s}$. Hence, if $R_{f} > \rvir$ the gas has expanded, doing
work in the process. $P_s$ is the surface pressure at $\rvir$, given
by $P_{s} = f_b P_{DM}(\rvir) = f_b \sigma^2_{DM}(\rvir)
\rho_{DM}(\rvir)$. The surface pressure of the gas at the virial
radius is thus equal to that of the dark matter multiplied by the
cosmic baryon fraction.  The second constraint used to solve the model
is that the total pressure of the gas at $R_f$ be equal to the gas
pressure at the virial radius of the cluster before polytropic
rearrangement, i.e., that $P_{tot}(R_f) = f_b P_{DM}(R_{vir})$.

One fundamental component of the model is the inclusion of heating of
the ICM via non-gravitational processes. This can occur via two
mechanisms: (1) through energy transfer from the dark matter to gas
during major mergers, and (2) via energy feedback from supernovae and
AGN outflows.

To account for the former, \citet{bode09} introduced the parameter
$\epsdm$ which controls the fraction of the total dark matter energy,
$|E_{DM}|$, transferred to the ICM during mergers \citep{pearce94,
  rasia04, lin06, mccarthy07}.  The total dark matter energy is simply
the sum of the total dark matter kinetic and potential energy
(calculated similarly to Equation \ref{eq:initenergy}, but setting
$r_* = 0$ and $f_b = 1$). From the hydrodynamical simulations of
\citet{mccarthy07}, B09 suggest $\epsdm = 0.05$.

Energy feedback from supernovae or AGNs is assumed, as a first
approximation, to be proportional to the total stellar mass of a
cluster and is determined by the parameter $\epsfb$. The approximate
value of this parameter is difficult to determine from hydrodynamical
simulations as methods of implementing AGN feedback have only recently
started to be investigated in earnest
\citep{sijacki07,dimatteo08,booth09,teyssier10}.  B09 demonstrated
that the feedback parameter can be calibrated by comparing the model
to low-redshift X-ray scaling relations, finding values in the range
$0 \leq \epsfb \leq 1.2\times 10^{-5}$ depending on assumptions of the
stellar mass of clusters and its mass dependence. Recently,
\citet{battaglia10} investigated the impact of AGN heating in
hydrodynamical simulations, finding an effective feedback efficiency
(which is roughly equivalent to our $\epsfb$) of $5\times 10^{-6}$ for
an $\m500 = 6.8 \times 10^{13} \hmsun$ cluster over the duration of
the simulation.

We note that, in terms of predicting the tSZ power spectrum, the
effect of $\epsfb$ and $\epsdm$ is somewhat degenerate.
Following B09, we take $\epsdm = 0.05$ as our fiducial
parameter. In the following section we compare our model using
different values of the feedback parameters $\epsfb$ and 
$\epsdm$ with direct X-ray observations of clusters. 
In Section \ref{sec:szpowerspec}, we investigate the impact of 
varying both parameters on the SZ power spectrum.

\subsubsection{Non-thermal Pressure}
\label{sec:nonthermpress}

Hydrodynamical simulations have demonstrated that a significant
fraction of the total energy of the ICM is contained within random gas
motions, which provide a significant non-thermal contribution to the
total pressure support \citep{evrard90, rasia04, kay04, dolag05,
  lau09}. This kinetic support is largely sourced by infalling and
merging structures, which can further generate turbulent gas motions
at the boundary between the bulk flow and the thermalized ICM
\citep{dolag05, kim07, vazza09}.  While the level of non-thermal
pressure is typically found to be small in the central regions of
clusters, it increases steadily with radius, becoming a significant
fraction of the total pressure at $\r500$ \citep{lau09}. The
simulation comparison study of \citet{frenk99} demonstrated that
different hydrodynamics codes predict a similar ratio of kinetic to
thermal gas energy in clusters.

To account for non-thermal pressure support in our model, we split the
total pressure into non-thermal and thermal components,
$P_{tot}(r) = P_{th}(r) + P_{nt}(r)$ and set the non-thermal pressure
fraction to be a power law with cluster-centric radius;
\begin{equation}
\frac{P_{nt}}{P_{tot}}(z) = \alpha(z) \left ( \frac{r}{\r500}\right)^{n_{nt}} \;\;,
\label{eq:nonthermalpress}
\end{equation}
where $n_{nt}$ determines the radial dependence of the non-thermal pressure fraction.

\begin{figure}
\includegraphics[scale = 0.26]{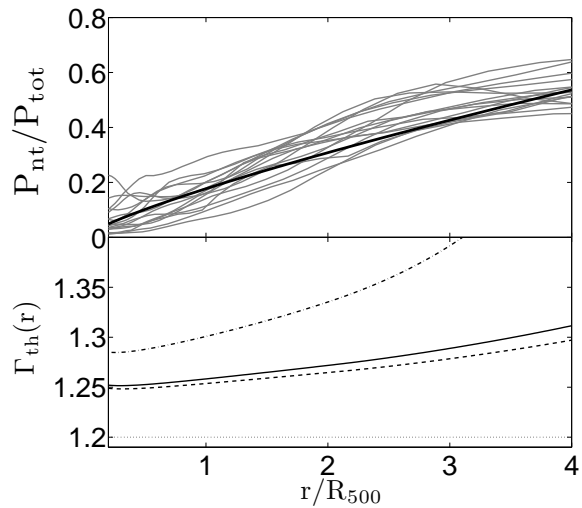}
\caption{({\bf Upper}) The ratio of non-thermal to total pressure as a
  function of radius for the 16 simulated clusters of \citet{lau09}
  (gray lines). The thick solid line is the best fit from Equation
  \ref{eq:nonthermalpress} and represents the fiducial non-thermal
  pressure fraction assumed in our model. ({\bf Lower}) The thermal
  polytropic index $\Gamma_{th}$ (see text) predicted by our gas model
  as a function of radius. The dotted, solid and dot-dashed black line
  represents $\alpha_0 = 0, 0.18$ and $0.3$, respectively. The dashed
  line represents the fiducial model $\alpha_0 = 0.18$, but with no
  star-formation or feedback.}
\label{fig:nonthermpress}
\end{figure}

We assume that the non-thermal pressure fraction varies with redshift
according to $\alpha(z) = \alpha_0 f(z)$, where $\alpha_0$ is the mean
ratio of non-thermal to total pressure at $\r500$ at $z = 0$, and
$f(z)$ is a monotonically increasing or decreasing function of
redshift. Since $P_{nt} / P_{tot} \leq 1$, Equation
\ref{eq:nonthermalpress} is valid for $\alpha(z) \leq (R_{\rm
  max}/\r500)^{-n_{nt}}$. As we describe below, the outermost radius
of the pressure profiles in our model is set to $4 R_{500}$, which
limits the maximum value to $\alpha(z) < 4^{-n_{nt}}$.  We adopt the
following form for the redshift evolution,
\begin{equation}
f(z) = {\rm min} [ (1+z)^\beta, (f_{max} - 1)\tanh (\beta z) +1 ] \;,
\label{eq:nonthermapress_zevo}
\end{equation}
where $f_{max} = 4^{-n_{nt}}/\alpha_0$ and $\beta$ is a free parameter
that determines the evolution rate. This form ensures that, at low
redshift ($z<1$), $\alpha(z)$ has a simple power-law dependence on
$(1+z)$, whereas at high redshift (or for large values of $\alpha_0$)
it smoothly asymptotes toward the maximum value of $4^{-n_{nt}}$. We
note that $\beta$ can be positive or negative. For the former, the
value of $\beta$ controls the redshift of transition from the
power law to the asymptotic behavior of $\alpha(z)$. For negative
values of $\beta$, $\alpha(z)$ declines toward zero with increasing
redshift.

We use the hydrodynamical simulations of \cite{nagai07} to find the
best-fit values of $\alpha_0$ and $n_{nt}$.  The upper panel of
Figure~\ref{fig:nonthermpress} shows the radial dependence of the
ratio of the non-thermal to total gas pressure fraction, $P_{nt} /
P_{tot}$, of 16 simulated clusters at $z=0$ (the same set of simulations
shown in Figure \ref{fig:phasediagram}).  Following \citet{lau09}, the
non-thermal pressure component (due to random gas motions) was
measured by determining the velocity dispersion of gas cells in radial
shells in the rest frame of the cluster. It is clear that the mean
non-thermal pressure fraction increases rapidly with radius, from
$<10\%$ at $r=0.1\r500$ to greater than $30\%$ at $r=2 \r500$. The
thick-solid line indicates the best-fit model with $\alpha_0 = 0.18
\pm 0.06$ and $n_{nt} =0.8 \pm 0.25$ (resulting in a maximum value of
$\alpha(z) = 0.33$).  We find no evidence of mass dependence on any of
the parameters. Furthermore, \citet{lau09} obtained similar results
for their non-radiative simulations (i.e., including no cooling or
star formation) outside the central regions of the clusters. This
indicates that the level of non-thermal pressure is not sensitive to
these processes. We note, however, that the preheating simulations of
\citet{stanek10} do suggest a lower level of kinetic energy in bulk
motions than in simulations without preheating.

We determine the redshift dependence by examining a high-redshift
output of the simulation.  At $z=1$, we find that the best-fit
$\alpha_0$ increases to 0.26, implying that $\beta \approx 0.5$. A
greater fraction of non-thermal pressure towards high redshift is
expected due to the increased rate of merger activity in the
$\Lambda$CDM model. \citet{stanek10} also observed an increasing level
of kinetic support with redshift in their `gravity-only'
smoothed-particle hydrodynamics simulation. However, the results of
their preheating simulation imply a decreasing ($\beta < 0$) level of
non-thermal pressure toward high redshift. The specific values of
$\alpha_0$ and $\beta$ predicted by simulations thus appear to be
somewhat dependent on the baryonic processes included.  We therefore
vary these parameters to investigate their impact on the tSZ
power spectrum. Further study is required to determine the dependence
of non-thermal pressure to cooling and heating effects in
hydrodynamical simulations.

We find that $n_{nt}$ decreases by approximately 15\% at $z = 1$
compared to the best-fit $z = 0$ value; however, as this change is less
than the 0.25 standard deviation on the redshift zero value, we
henceforth hold $n_{nt}$ constant with redshift. Our fiducial model is
therefore $\alpha_0 = 0.18$, $n_{nt}=0.8$ and $\beta=0.5$.  We
investigate the impact of varying the overall normalization of the
profile $\alpha_0$ and $\beta$ on the electron pressure profiles of
clusters and the tSZ power spectrum, while fixing $n_{nt}$ to the
fiducial value for the remainder of the paper. In a future work, we
intend to investigate in more detail the mass and redshift dependence
of non-thermal pressure profiles in much larger samples of simulated
groups and clusters.

The radial thermal pressure profile given by our model is
\beq
P_{th}(r) = P_o \theta(r)^{n+1} \left[ 1- \alpha(z) \left ( \frac{r}{\r500}\right)^{0.8}\right] \;.
\label{eq:pressprof}
\eeq
The gas density profile remains as in Equation \ref{eq:model_press_den}. The radial 
temperature profile is then given by
\beq
k T(r) = \mu m_p \frac{P_0}{\rho_0} \theta(r) \left[ 1- \alpha(z) \left ( \frac{r}{\r500}\right)^{0.8}\right] \;,
\label{eq:Tprof}
\eeq
where $\mu$ is the mean molecular weight and $m_p$ is the proton mass.  

One consequence of including of non-thermal pressure support in our
model is that the adiabatic index of the thermal component of the gas,
given by $\Gamma_{th} = \log(P_{th}/P_0)/\log(\rho_g/\rho_0)$, varies
as a function of cluster radius. X-ray observations of clusters
\citep{vikhlinin06, deGrandi02} as well as hydrodynamical simulations
\citep{battaglia10} indicate that the ICM thermal pressure and density
profiles do not adhere to a polytrope with constant index.  The lower
panel in Figure \ref{fig:nonthermpress} shows $\Gamma_{th}$ as a
function of radius produced by our model for a $\m500 = 3\times
10^{14} \hmsun$ cluster. The solid lines shows the results for
$\alpha_0 = 0.18$, the dot-dashed line for $\alpha_0 = 0.30$ and the
dotted line for $\alpha_0 = 0$. The dashed line gives the results for
$\alpha_0 = 0.18$, but assuming $\epsfb = \epsdm = M_* = 0$. In the
zero non-thermal pressure case, we obtain $\Gamma_{th} = \Gamma =
1.2$, as expected. Setting $\alpha = 0.18$ (0.3) results in a $\approx
5$ $(8)\%$ increase in $\Gamma_{th}$, which also increases with
radius.  Turning off star formation and feedback in our model reduces
this radial dependence slightly.

\section{Comparison with X-ray observations}
\label{sec:calibration}

A fundamental test of our model is to compare with the global
properties and radial profiles of clusters obtained from X-ray
observations. The main aim is to evaluate the range in which the two
principal parameters in this model, the feedback parameter $\epsfb$
and the non-thermal pressure support parameter $\alpha_0$, reproduce
observed cluster properties. We focus specifically on the
$\m500-\fgas$ relation measured from the cluster samples of
\citet{vikhlinin06} and \citet{sun09} (where the gas fraction $\fgas$
is defined as the gas mass divided by total mass within $\r500$), and
the `universal' electron pressure profile measured by \citet{arnaud10}
from the REXCESS cluster sample \citep{bohringer07, pratt09}. The
latter provides an important test as the radial thermal pressure
profile is the cluster property that determines the SZ power
spectrum. We have chosen to use the $\m500-\fgas$ relation as we find
the gas fraction in our model to be particularly sensitive to the
precise values of $\epsfb$.

\subsection{Hydrostatic Mass Estimates}

One of the principal techniques for measuring cluster masses is to measure 
the radial gas density and temperature profiles and solve the equation of
hydrostatic equilibrium to derive the total mass profile, 
\beq 
M(<r) = \frac{-r^2}{G\rho_{g}}\frac{dP_{tot}}{dr} \;,
\label{eq:hse}
\eeq \citep{sarazin86, evrard96}.  However, direct observations of
clusters only currently probe the thermal pressure profile and so miss
the significant contribution of non-thermal pressure in the ICM. This
can result in a systematic underestimation of $\m500$ by $10\%-20\%$
\citep[e.g.,][]{rasia04, nagai07b, piffaretti08, meneghetti10}. Hence
one must take care when comparing simulation or model scaling
relations with observations that utilize this technique.

\citet{vikhlinin06} and \citet{sun09} use this method to measure the
masses of the groups and clusters in their sample. \citet{arnaud10}
determine the mass dependence of the amplitude of the pressure
profiles of the clusters in their sample using the $Y_X- \m500$
relation calibrated by \citet{vikhlinin06}.  Therefore, in order to
compare our model with the observational data, we use the hydrostatic
mass predicted by our model $\mhse$ rather than the true $\m500$.

\begin{figure}
\includegraphics[scale = 0.26]{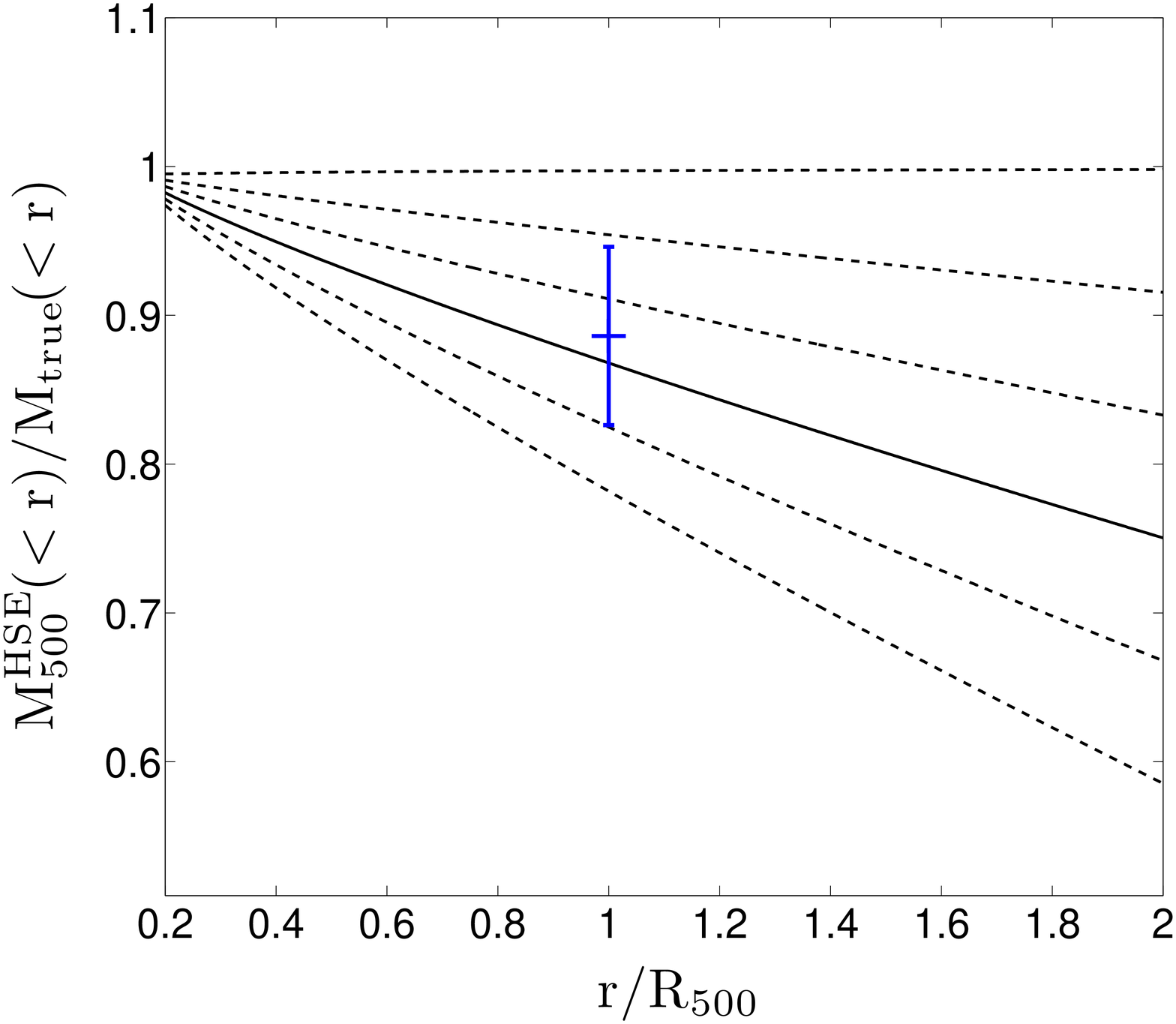}
\caption{Ratio of hydrostatic to true mass measured within spheres of
  radius $r/\r500$ for a cluster of $\m500 = 3\times 10^{14}
  \hmsun$. We vary the values of the non-thermal pressure support
  parameter, $0 \leq \alpha_0 \leq 0.3$ (top to bottom) in steps of
  0.06. The solid line represents the fiducial value $\alpha_0 =
  0.18$. The blue point represents the mean $\mhse/\m500$ measured
  from the simulated clusters of \citet{lau09}, with the error bar
  representing the error on the mean.}
\label{fig:mhse}
\end{figure}

Figure \ref{fig:mhse} shows the ratio of hydrostatic to true mass
obtained from our model, varying the value of the non-thermal pressure
parameter, $\alpha_0$ from zero (top line) to 0.3 (bottom line) in
steps of 0.06. The thick solid line represents our fiducial value of
$\alpha = 0.18$. The blue point represents the mean $\mhse/\m500$
measured from the \citet{lau09} simulated clusters, with the error bar
representing the error on the mean. 

For our fiducial model, $\mhse$ underestimates the true mass $\m500$
by a factor of $0.87$ at $r=\r500$, increasing to $0.75$ at
$r=2\r500$. Setting $\alpha_0 = 0.3$ increases this to $0.78$ and
$0.59$ at these radii, respectively.  We find that changing the
feedback parameter $\epsfb$ has only a small effect on the
hydrostatic mass estimate.

Note that the dependence of $\mhse$ on $\alpha_0$ presents a
complication in comparing the model with observations as holding
$\mhse$ constant requires varying the true mass $\m500$ as we vary the
model parameters. This makes it difficult to determine whether changes
in predicted properties are due to changes in the true mass or model
parameters. To overcome this, in the remainder of this section we fix
$\mhse = 0.87\m500$ (the ratio obtained from our fiducial value of
$\alpha_0$) for all model realizations.

\subsection{$\mhse-\fgas$ relation}

\begin{figure}
\includegraphics[scale = 0.26]{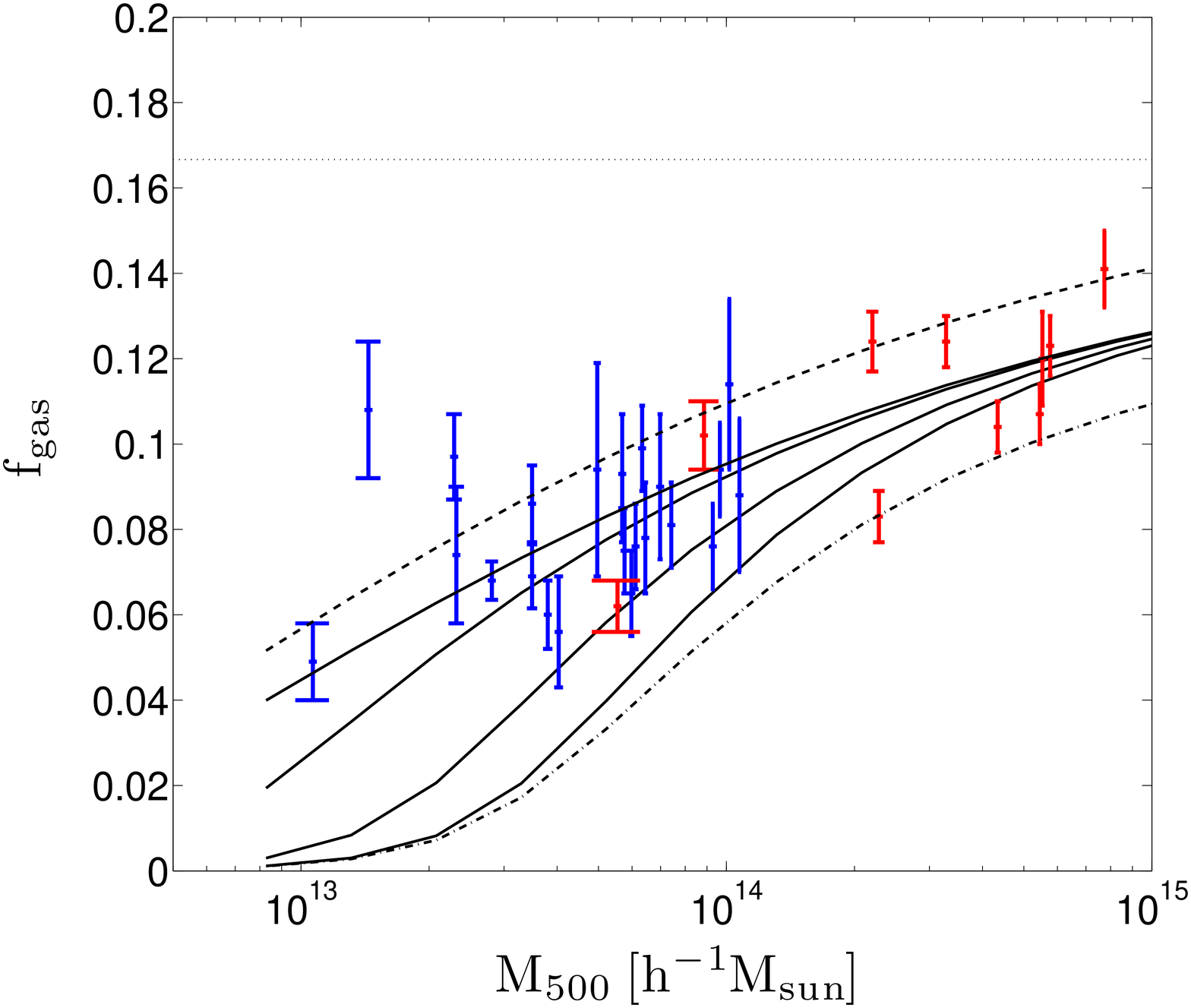}
\caption{Comparison between the observed and model $\mhse - \fgas$
  relation for increasing values of the feedback parameter $\epsfb =
  10^{-7}, 10^{-6}, 5\times 10^{-6}$, and $10^{-5}$ (solid lines, from
  top to bottom). Points with error bars represent individual cluster
  observations from \citet{sun09} (blue) and \citet{vikhlinin09}
  (red). The dashed line represents the minimal feedback model
  ($\epsfb = \epsdm = 0.0$; i.e., star-formation only). The dot-dashed
  line is the maximal feedback model with $\epsfb = 10^{-5}$ and
  $\epsdm = 0.1$. The horizontal dotted line represents the universal
  baryon fraction.}
\label{fig:mfgas}
\end{figure}

In Figure \ref{fig:mfgas}, we compare the $\mhse-\fgas$ relation
predicted by our model for different values of $\epsfb$ against the
X-ray observations. The red points are from \citet{vikhlinin06},
whereas the blue points are from \citet{sun09}. The solid lines represent
the results from our ICM gas model with (from top to bottom) $\epsfb =
10^{-7}, 10^{-6}, 5\times10^{-6}$, and $10^{-5}$. For comparison, we
also plot results for $\epsfb = \epsdm = 0$ (dashed line) and a
maximal feedback model with $\epsdm = 0.1$, $\epsfb = 10^{-5}$
(dot-dashed line).  In this plot, $\alpha_0$ is kept fixed at the fiducial
value of 0.18 (note that as we assume a polytropic equation of state
between gas density and {\it total} pressure, changing the non-thermal
pressure fraction does not change $\fgas$). The dotted line indicates
the cosmic baryon fraction, $\Omega_b / \Omega_M$. We have verified
that our model predicts $\fgas \approx \Omega_b / \Omega_M$ in the
absence of star formation and feedback.

It is clear that for high-mass clusters varying the feedback parameter
$\epsfb$ (solid lines) produces little change in the gas fraction. The
energy added to the ICM is small compared to the total binding energy
for these clusters and thus does not disrupt the gas distribution
significantly. However, the impact of energy feedback becomes
increasingly important towards lower mass. As cluster mass decreases
the feedback energy becomes a larger fraction of the total binding
energy. In these systems, feedback has the effect of `inflating' the
gas distribution, thus reducing the baryon fraction within a fixed
radius.

Setting $\epsdm = 0$ (dashed line) demonstrates that the dark matter
energy transfer parameter has a constant effect with mass, reducing
$\fgas$ by $0.01-0.02$ across the entire mass range. This is
because the energy transferred to the gas from the dark matter is
proportional to the binding energy of the cluster, and so has a more
significant effect on the high mass clusters than $\epsfb$.

There is generally a large amount of intrinsic scatter in the observed
sample, especially at the high-mass end where the error bars on
individual estimates of $\fgas$ are smaller. Although the
\citet{vikhlinin06} clusters were selected for their relaxed
morphology, the large scatter may indicate a range of merger histories
within the sample. It may also reflect the large ($\approx 50\%$)
intrinsic scatter observed in the mean stellar mass -- total mass
relation \citep{giodini09, gonzalez07}. Nevertheless, the upper and
lower model profiles plotted in Figure \ref{fig:mfgas} bracket the
observed data points. We therefore assume this range in $\epsfb$
($0$--$10^{-5}$) and $\epsdm$ (0--0.1) when evaluating the modeling
uncertainty on the tSZ power spectrum.

\subsection{Electron Pressure Profiles}
\label{sec:pressprof}

\begin{figure*}
\centering
\subfigure{
\includegraphics[scale = 0.27]{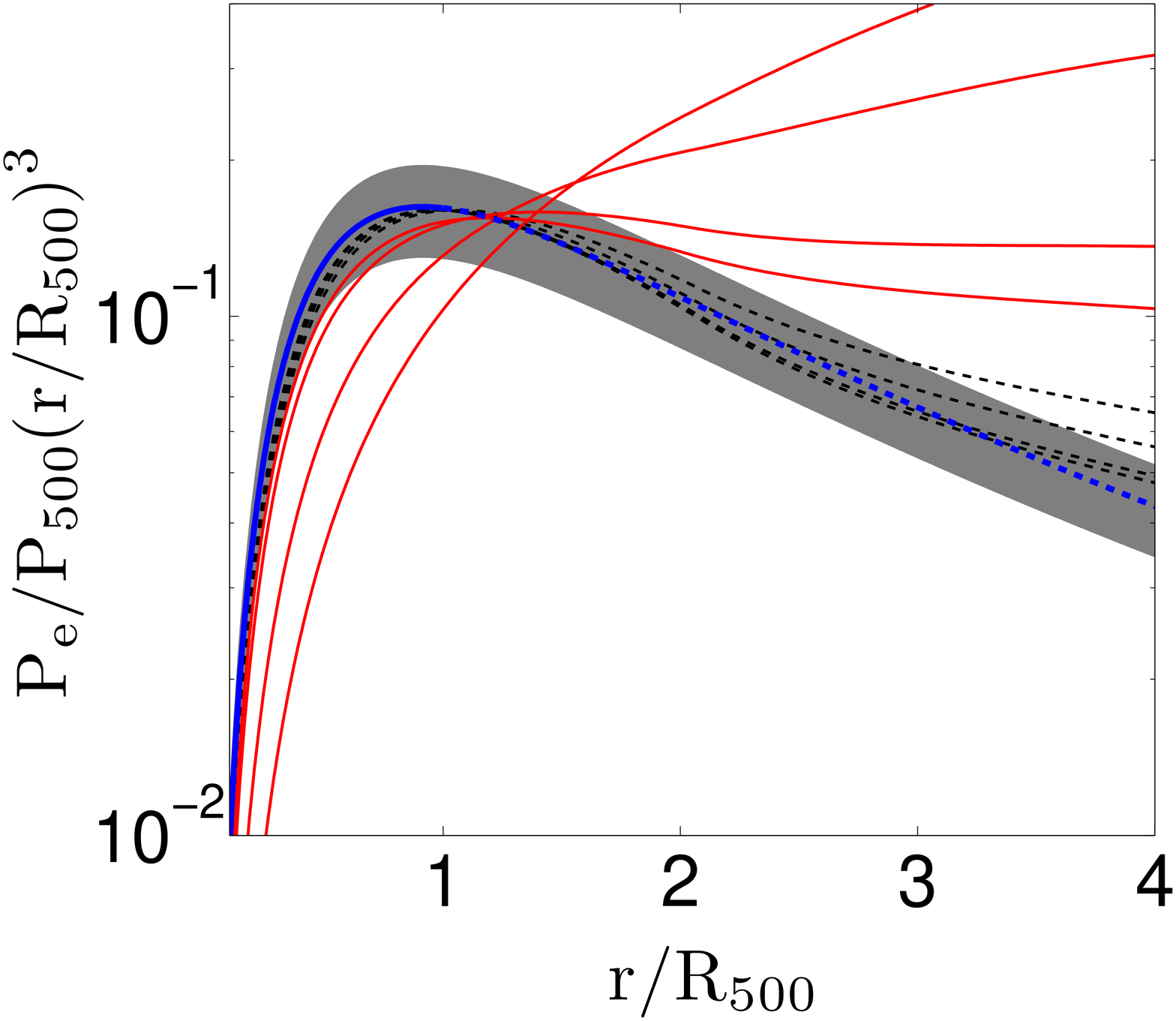}
}
\subfigure{
\includegraphics[scale = 0.27]{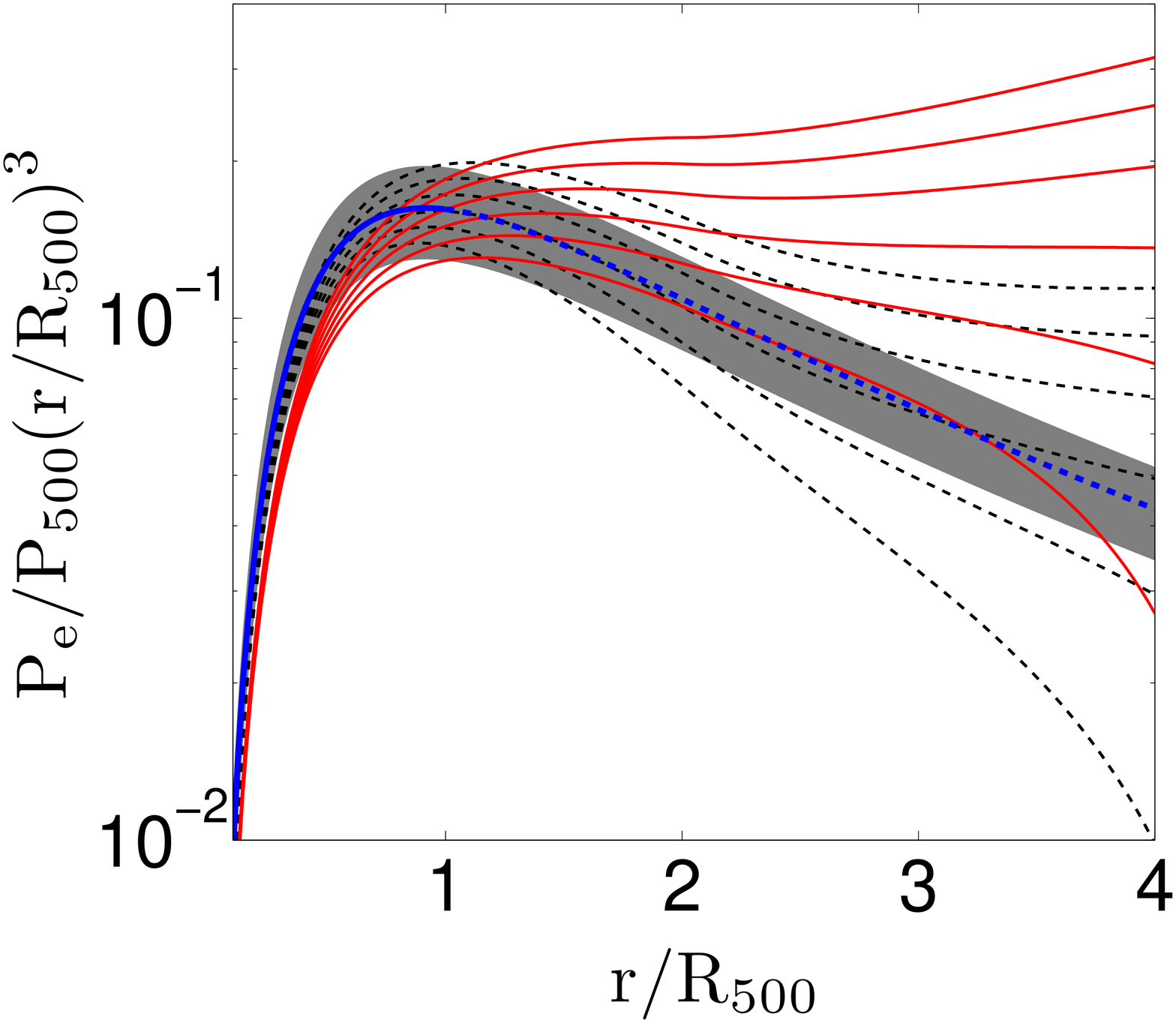}
}
\caption{({\bf Left}) Electron pressure profiles for clusters of mass
  $\mhse = 3\times10^{14} \hmsun$ (black dashed lines) and $\mhse =
  3\times 10^{13} \hmsun$ (red lines), using feedback parameters
  values of $\epsfb = 10^{-7}, 10^{-6}, 5\times 10^{-6}$, and $10^{-5}$
  (from bottom to top at $2\r500$). The blue line represents the A10
  pressure profile, with the shaded region denoting the observed
  20\% dispersion around this profile.  ({\bf Right}) Same as the upper
  panel but for increasing values of the non-thermal pressure support
  parameter $0 \leq \alpha_0 \leq 0.30$ in steps of 0.06 (from top to
  bottom)}.
\label{fig:Peprofile}
\end{figure*}

The two key components required to calculate the tSZ power spectrum
are the halo mass function and the projected radial electron pressure
profiles for clusters over a wide range of mass and redshift. An
important test is therefore to compare directly the three-dimensional pressure
profiles of our model against those measured from X-ray
observations. 

\citet[][henceforth A10]{arnaud10} measured the electron pressure
profile for intermediate to high-mass ($10^{14} < \mhse/\msun <
10^{15}$), low redshift ($z<0.2$) clusters in the REXCESS sample
\citep{bohringer07, pratt09}. The scaling of the amplitude of the
profile with cluster mass at $\r500$ was found to be $P_e \propto
M^{0.69\pm 0.16}$ and thus in agreement with self-similar expectations
($P_{500} \propto M^{2/3}$, where $P_{500}$ is the characteristic
pressure in the self-similar model as described in Appendix A of
A10). Due to the low-redshift nature of the REXCESS sample, A10 were
not able to constrain significantly the redshift evolution of the
sample, assuming that it follows the self-similar form, $P_{500}
\propto h(z)^{8/3}$.

Having rescaled the profiles to remove any mass dependence there was
found to be little dispersion (approximately 20\%) around the mean
profile outside of the core region, $r > 0.2\r500$. Within $0.2
\r500$, A10 observed a much larger dispersion in the measured pressure
profiles, with the shape of the profile related to the dynamical state
of the cluster.  A10 found there to be a good agreement between the
observed profiles and those measured from the hydrodynamical
simulations of \citet{borgani04, nagai07} and \citet{piffaretti08}.
As the observations of A10 extend only out to $\r500$, the simulation
data were used to extend the best-fit profile out to $4\r500$.

\subsubsection{Impact of Energy Feedback}

In the left panel of Figure \ref{fig:Peprofile}, we compare the
electron pressure profiles for our gas model over a range of values of
$\epsfb$ against the A10 profile. We plot the pressure profiles scaled
as $P_e(r) / P_{500} (r/\r500)^3$ to allow a clear comparison of the
profiles in the outer regions, which contribute significantly to the
tSZ power spectrum at the angular scales of interest. The solid blue
line represents the P09 profile within $\r500$, the radius within
which it was observed. The dashed blue line represents the region in
which the profile was determined from simulations rather than
observations.  The shaded region denotes the $20\%$
dispersion observed by A10 around the mean profile. We plot $P_e(r)$
for two different masses, $\mhse = 3 \times 10^{14} \hmsun$ (black
dashed lines) and $\mhse = 3\times10^{13} \hmsun$ (red solid), both at
$z = 0.1$. We note that the A10 profile was measured for clusters of
mass $\mhse > 7\times10^{13} \hmsun$ and so the lower of the plotted
masses represents an extrapolation of the mass dependence of this
profile. As in Figure \ref{fig:mfgas}, we plot model profiles for
$\epsfb = 10^{-7}, 10^{-6} , 5\times10^{-6}$ and $10^{-5}$. $\alpha_0$
is fixed at the fiducial value (0.18).

For $\mhse = 3\times10^{14}\hmsun$ (black dashed lines), varying the
feedback parameter has a very small effect on the pressure profile,
especially within $\r500$. As noted above, for higher mass clusters
the feedback energy is a small fraction of the total binding energy
and therefore does not strongly influence the gas density and
temperature distribution.  This is not the case for lower mass
clusters. For $\mhse = 3\times10^{13} \hmsun$, increasing the feedback
lowers the overall electron pressure within $2\r500$, with the effect
strongest in the central regions of the cluster.  As demonstrated in
Figure \ref{fig:mfgas}, increasing the feedback parameter has the
effect of inflating the gas mass distribution, reducing the gas
density within $\r500$ and increasing it at larger radii. Although
increasing the feedback parameter produces a small rise in gas
temperature, this is outweighed by the decrease in the gas density,
thus lowering the overall thermal pressure within $\r500$. We note
that at larger radii ($r> 1.2\r500$), the electron pressure increases
with $\epsfb$.

For high-mass clusters we find that all values of $\epsfb$ produce an
excellent match to the A10 pressure profile in the range $0.1 \leq
r/\r500 \leq 2$. For $\m500 = 3\times10^{13} \hmsun$, a
significantly lower mass than any of the observed clusters in the A10
sample, the amplitude of the pressure profile is consistently below
that of the A10 profile within $\r500$, but above at larger radii.

\subsubsection{Impact of Non-thermal Pressure Support}

In the right panel of Figure \ref{fig:Peprofile}, we show the
dependence of the gas model pressure profile on the non-thermal
pressure support parameter $\alpha_0$. We vary $\alpha_0$ in range $ 0
\leq \alpha_0 \leq 0.30$ in steps of 0.06 (from top to bottom at
$\r500$). As in the upper panel, the two sets of curves represent
clusters of mass $\mhse = 3\times10^{13}$ and $3\times10^{14}\hmsun$,
respectively. The feedback parameter is fixed at $10^{-6}$, which we
henceforth take as our fiducial value.

As the fraction of non-thermal pressure support is increased, the
profiles become steeper, significantly reducing the thermal pressure
in the outer regions. This is entirely expected from Equation
\ref{eq:pressprof}. For $\mhse = 3\times 10^{14} \hmsun$, setting
$\alpha_0 = 0.18$ produces a good fit to the A10 profile all the way
out to $3 \r500$. In general, all the values of $\alpha_0$ explored
produce a pressure profile that, for $r < \r500$, lies within the 20\%
dispersion observed around the A10 profile. For $\m500 = 3\times
10^{13} \hmsun$ the profiles tend to lie below the A10 profile within
$\r500$, but predict a higher pressure beyond this radius.

\subsubsection{Mass and Redshift Dependence}

At $r=\r500$, our fiducial model ($\epsfb = 10^{-6}$ and $\alpha_0 =
0.18$) produces a mass scaling of $P_e(\r500) \propto
M^{0.75\pm0.01}$. At $0.5 \r500$, this steepens to
$M^{0.79\pm0.01}$. This steepening of the mass scaling at smaller
radii was also observed by A10, who found a scaling of
$M^{0.69\pm0.16}$ and $M^{0.78\pm0.16}$ at $r=\r500$ and $0.5 \r500$,
respectively\footnote{We assume the same error on the mass scaling at
  $0.5\r500$ as was measured at $\r500$ for the A10 profile.}.

\begin{figure}
\includegraphics[scale = 0.26]{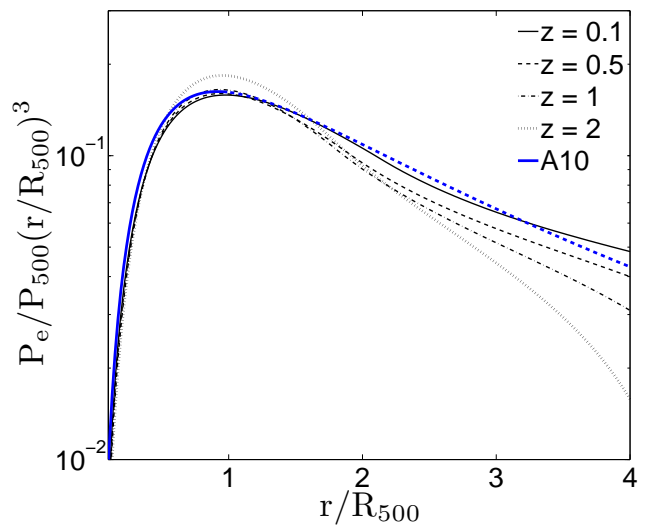}
\caption{Pressure profiles for model clusters at different redshifts,
  where $z = 0, 0.5, 1,$ and $2$ (solid, dashed, dot-dashed, dotted
  and thin solid lines, respectively). The blue line is the observed
  pressure profile of A10.  The cluster mass is fixed at $\mhse =
  3\times10^{14} \hmsun$.}
\label{fig:Pe_redshifts}
\end{figure}

In Figure \ref{fig:Pe_redshifts}, we plot the redshift evolution of the
pressure profile where the solid, dashed, dot-dashed, and dotted lines
correspond to $z = 0, 0.5, 1$ and $2$, respectively. The blue lines
represent the A10 profiles at each redshift, for which the
normalization is assumed to scale self-similarly, i.e., $P_{500}
\propto h(z)^{2.67}$.

At $r \approx 1.5\r500$ our model reproduces the self-similar scaling
of the A10 profile, but scales differently with redshift at smaller
and larger radii. As redshift increases, the profile becomes more
centrally concentrated, with the pressure increasing with respect to
the A10 profile for $r \leq 1.5\r500$, but decreasing at larger
radii. At $r=0.5, 1$ and $2\r500$ we find that the profile scales as
$\propto h(z)^{2.71\pm0.04}$, $h(z)^{2.79\pm0.05}$ and $h(z)^{2.57\pm0.06}$.

There are three components of our fiducial model that evolve with
redshift at fixed halo mass: the stellar mass fraction $M_* / \m500$
(and thus the feedback energy), halo concentration -- both of which
decrease towards increasing redshift -- and the non-thermal pressure
parameter $\alpha$, which initially increases as $(1+z)^{0.5}$ but
asymptotes toward an upper limit of $\alpha(z)= 0.33$ at high
redshift (Section \ref{sec:nonthermpress}). While the increasing
non-thermal pressure support produces a negative redshift evolution,
this is compensated to some extent by the decreasing concentration.
At fixed mass, as halo concentration decreases so does the central
gravitational potential, reducing the pressure in the inner regions as
well as increasing it at larger radii. We find that removing the
redshift evolution of the concentration parameter (Equation
\ref{eq:massconc}) results in a significant weakening of the redshift
evolution relative to self-similar, $P_e(2\r500) \propto
h(z)^{1.84\pm0.05}$. However, the evolving stellar mass fraction does not
have a strong effect on the redshift scaling of the pressure profiles.

\section{Sunyaev-Zel'dovich Angular Power Spectrum}
\label{sec:szpowerspec}

\begin{figure*}
\plotone{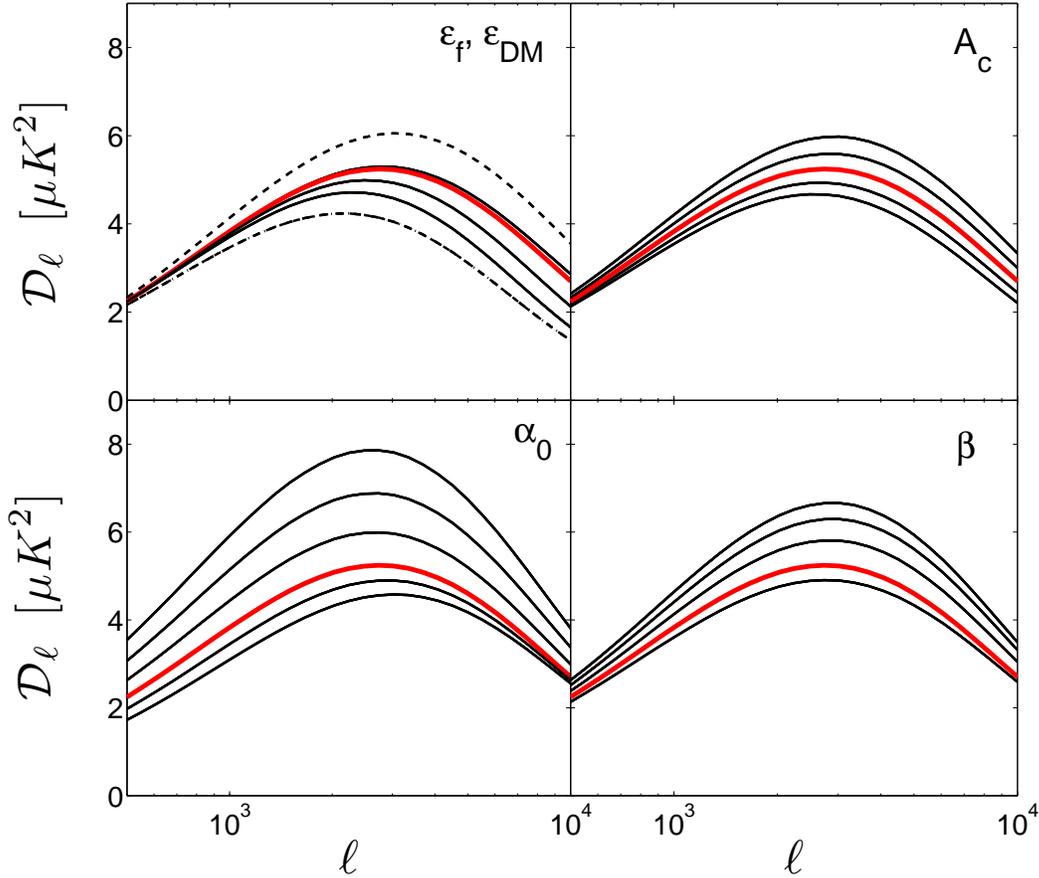}
\caption{({\bf top, left}) The tSZ power spectrum for different values
  of $\epsfb$ and $\epsdm$. At $\ell = 10,000$, the values of $\epsfb$
  are (from top to bottom) $10^{-7}, 10^{-6}, 5\times 10^{-6}$ and
  $10^{-5}$. The dashed line denotes $\epsdm = \epsfb = 0$. The
  dot-dashed line indicates a model with $\epsfb = 10^{-5}$ and
  $\epsdm = 0.1$.  ({\bf top, right}) tSZ power spectrum for different
  normalizations of the halo concentration-mass relation. At $\ell =
  10,000$, $A_C=0.8, 0.9, 1.0, 1.1, 1.2$. (from bottom to top) ({\bf
    bottom, left}) tSZ power spectrum for increasing values of
  $\alpha_0$, where, from top to bottom, $0 \leq \alpha_0 \leq 0.30$ in
  steps of $0.06$. ({\bf bottom, right}) Varying the value of the
  non-thermal pressure support redshift evolution parameter $\beta$ in
  the range $-1 \leq \beta \leq 1$ (from top to bottom) in steps of
  0.5 . In each case, the thick red line represents our fiducial
  model.}
\label{fig:powerspectra}
\end{figure*}

The primary goal of this work is to investigate the variations in the
predicted tSZ power spectrum caused by uncertainties in ICM physics,
in this case parameterized by the energy feedback and non-thermal
pressure support parameters described in Sections \ref{sec:feedback}
and \ref{sec:nonthermpress}. In the previous section we demonstrated
the effect of varying these parameters on the pressure profile for
individual clusters, comparing with results from high-quality X-ray
cluster observations. We now repeat this exercise for the tSZ power
spectrum. We also compare our fiducial model with previous simulations
or models and the recent SPT observations of the tSZ power spectrum.

\subsection{Impact of Cluster Physics}

\begin{table}[t]
\begin{center}
\begin{tabular}{|c | c | c | c | c | c |}
\hline
Value & $\epsilon_{\rm f}$ & $\epsilon_{\rm DM}$ & $\alpha_0$  & $\beta$ & $A_C$ \\
\hline
Fiducial & $10^{-6}$ & 0.05 & 0.18 & 0.5 & 1.0 \\
\hline
Min. & 0 & 0 & 0 & -1 & 0.8\\ 
\hline
Max. & $10^{-5}$ & 0.1 & 0.3 & 1 & 1.2 \\
\hline
\end{tabular}
\caption{Range of Gas Model Parameters.}
\label{tab:paramrange}
\end{center} 
\end{table}

Figure \ref{fig:powerspectra} illustrates the tSZ power spectrum
obtained while varying the energy feedback parameters ($\epsfb$ and
$\epsdm$), the normalization of the halo mass-concentration relation
($A_C$), the non-thermal pressure support parameter ($\alpha_0$) and
its redshift dependence ($\beta$). In each plot, the thick red line
denotes the power spectrum obtained from our fiducial model
parameters. Table \ref{tab:paramrange} summarizes the fiducial,
maximum and minimum of each parameter explored. The range within which
we vary $\alpha_0$, $\epsfb$ and $\epsdm$ are chosen such that our
model brackets the observations described in the previous section.

We plot the band powers in terms of $\dell = \ell(\ell+1){\rm C_\ell}/
(2\pi)$ in units of $\mu K^2$ at 147 GHz (i.e., $f(x_\nu) = -1$ in
Equation \ref{eq:tsz_powerspec}).  Note that we truncate our model
profiles at $2 \rvir (\approx 4\r500)$. This radius was chosen to
roughly coincide with that of the accretion shock that is observed in
hydrodynamical simulations \citep[e.g.,][]{molnar09}. We do not expect
our model to be valid beyond the shock radius. Furthermore, the
contribution of the low-density gas outside of groups and clusters to
the tSZ power spectrum is expected to be small
\citep{hallman09, trac10}. Nevertheless, by doubling (tripling) the
outermost radius we find an average increase of only $6\% \; (7\%)$ in
tSZ power. 

In the top-left panel, we plot the tSZ power spectrum varying the
feedback parameter between $\epsfb = 10^{-7}, 10^{-6}, 5\times
10^{-6}$ and $10^{-5}$ (solid lines from top to bottom at $\ell =
10,000$). The dashed line represents $\epsfb = \epsdm = 0$ and the
dot-dashed line $\epsfb = 10^{-5}, \epsdm = 0.1$. Increasing $\epsfb$
and $\epsdm$ has the effect of suppressing power at small angular
scales, causing $\dell$ to peak at increasingly large scales. For
example, for $\epsfb = \epsdm = 0$, $\dell$ peaks at $\ell = 3000$,
while for the largest feedback values probed, $\dell$ peaks at $\ell =
2000$. Similar results have recently been reported by
\citet{battaglia10}, who investigate the impact of AGN feedback in
hydrodynamical simulations on the SZ power spectrum. The suppression
of small-scale power is caused by the inflating effect of energy
feedback flattening the pressure profiles in the inner regions of
clusters. As demonstrated in Figure \ref{fig:Peprofile}, increasing
feedback has the strongest effect on lower mass clusters
$M<10^{14}\hmsun$, which principally contribute to the power spectrum
on small angular scales.

In the lower-left panel of Figure \ref{fig:powerspectra}, we show the
tSZ power spectrum for increasing values of the non-thermal pressure
parameter, $\alpha_0$. We increase $\alpha_0$ in steps of 0.06 between
$0$ and $0.30$ (from top to bottom at $\ell = 1000$), while holding
the other gas model parameters at their fiducial value. 

Increasing $\alpha_0$ results in a significant decrease in power,
particularly at large angular scales; at $\ell = 3000$, the full range
in $\alpha_0$ encompasses more than a factor of 2 in power. In Section
\ref{sec:pressprof} we demonstrated that increasing the non-thermal
pressure parameter significantly decreases the thermal pressure at
large radii. This drives the large reduction in tSZ power,
particularly at low $\ell$.  We note that for higher values of
$\alpha_0$, $\alpha(z)$ rapidly reaches the maximum value of $0.33$
for the fiducial value of $\beta = 0.5$. The effect of increasing
$\alpha_0$ therefore begins to saturate for $\alpha_0 > 0.18$. This is
particularly noticeable on small scales ($\ell \geq 5000$), where high
redshift groups and clusters contribute significantly. It is evident
that determining the level of non-thermal pressure support in groups
and clusters, from all sources, is important in predicting the
amplitude of their tSZ signal.

In the lower-right panel of Figure \ref{fig:powerspectra}, we plot
(from top to bottom) the tSZ power spectrum for $\beta =
-1,-0.5,0,0.5$ and $1$. We hold $\alpha_0$ fixed at 0.18. It is clear
that the largest absolute change in power is at $\ell=2500$, although
the fractional change is similar across the full range of angular
scales plotted. This is in contrast to varying $\alpha_0$, which
produces an increasing effect toward lower $\ell$. The reason for
this is that varying $\beta$ predominantly affects high redshift
objects, which contribute to the tSZ power spectrum at smaller angular
scales than lower redshift objects of the same mass. However,
$\alpha_0$ primarily influences the pressure profiles at large radii
(i.e., larger angular scales). Hence, there is a trade-off with the
largest change in power being found at intermediate scales.

In the upper-right panel we plot the tSZ power spectrum while varying
the normalization of the mass-concentration relation around the
\citet{duffy08} value. This is achieved by setting $A_C = 0.8, 0.9,
1.0, 1.1$ and $1.2$ (from bottom to top at $\ell = 10,000$), where
$A_C$ is defined in Equation \ref{eq:massconc}. The results show that
increasing halo concentration boosts the SZ power at small angular
scales.  For a given mass, higher concentration halos have a deeper
central potential, thus increasing the pressure required for the inner
regions of clusters to maintain gas in hydrostatic equilibrium. This
steepens the pressure profiles in the inner regions of clusters,
resulting in the observed increase in power at small angular
scales. It is interesting to note that the impact of varying $A_C$ on
the tSZ power spectrum is somewhat similar to that of varying the
feedback parameter $\epsfb$.

\subsection{Comparisons with Simulations}

\begin{figure*}
\plotone{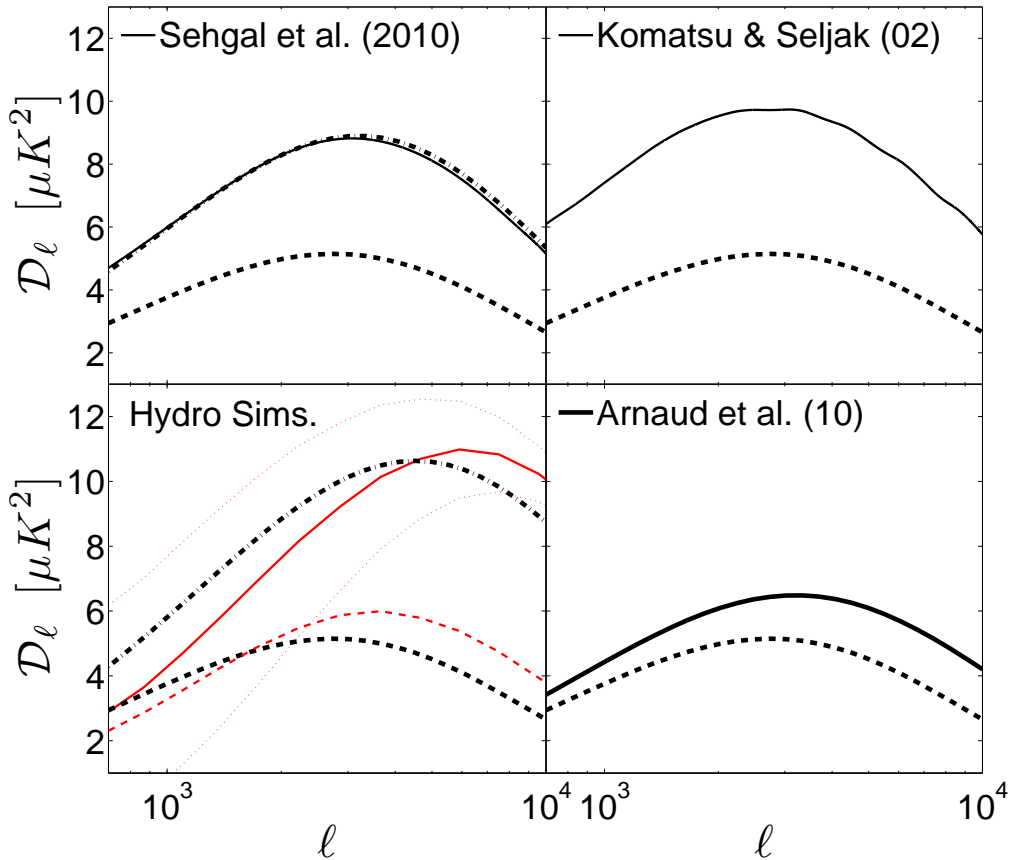}
\caption{Comparison with simulations and other models (solid
  lines). In each panel, the black dashed line is our fiducial model
  and the dash-dot line our model having modified the input parameters
  to reflect the level of physics incorporated in the simulation
  (where applicable). In the lower-left panel, the red lines show the
  results of the hydrodynamical simulations of
  \citet{battaglia10}. All results are plotted at 147 GHz and are
  scaled to $\sigma_8 = 0.8$ and $\Omega_M = 0.264$.}
\label{fig:powerspectra_sim}
\end{figure*}

In Figure \ref{fig:powerspectra_sim}, we compare our model to other
recent models or simulations. When comparing with simulations, we plot
both our fiducial model (dashed lines) and a second realization in
which we adjust the model parameters to match the physics assumed in
each simulation (dot-dashed lines), for example, by turning off
star formation and energy feedback. This enables us to examine the
impact of second-order cluster properties (such as morphology and
substructure) that are not accounted for in our halo-model approach.

In the top-left panel we compare our model with the simulations of
\citet{sehgal10}, who calculated the SZ power spectrum using simulated
SZ sky maps generated by applying the semi-analytic model of
\citet{bode09} to halos identified in an N-body lightcone
simulation. This approach thus includes the effects of variations in
halo density profiles, scatter in the mass-concentration relation and
halo morphology (although the requirement of hydrostatic equilibrium
will suppress the impact of substructure). The key difference between
the model of \citet{bode09} and that described here is our inclusion
of radially dependent non-thermal pressure support. The dot-dashed
line demonstrates that we are able to reproduce almost exactly the
\citet{sehgal10} tSZ power spectrum when we set $\alpha_0 = 0$.  This
implies that variations in dark matter halo structural properties, do
not strongly affect the tSZ power spectrum at the angular scales
investigated here.  Comparing the \citet{sehgal10} prediction with our
fiducial model highlights the importance of incorporating non-thermal
pressure support in ICM gas models.

In the top-right panel, we compare our fiducial power spectrum with
that predicted by the model of \citet{komatsu02}.  This model also
assumes that the ICM is in hydrostatic equilibrium, but does not
include star formation or any non-gravitational heating
mechanisms. While it is difficult to isolate the exact cause of the
difference between the two profiles, \citet{battaglia10} have recently
demonstrated the \citet{komatsu02} model significantly
over-predicts the thermal gas pressure at large radii in comparison to
observations, which is consistent with the factor of 2 increase in
power compared to our model. 

In the lower-left panel we compare our model with SZ power spectra
measured from recent hydrodynamical simulations. The red-solid line
shows the non-radiative, GADGET-2 simulation of \citet{battaglia10},
and the red-dashed line shows the result of a rerun of this
simulation including cooling, star formation and AGN feedback. The
dotted lines show the expected sample variance around the
non-radiative simulation given the simulated map size
\citep[$1.6^{\circ}\times 1.6^{\circ}$, see also][]{shaw09}. The black
dashed line shows our fiducial model. For comparison, we also plot our
model prediction having set the stellar mass and feedback energy to
zero (black, dash-dotted line).

The simulation without star formation, cooling and feedback predicts a
higher tSZ signal than the one that includes these
processes. On small scales, this is principally due to the suppression
of the SZ signal in the central regions of lower mass structures as
gas is expelled by AGNs \citep{battaglia10}. On larger scales the
difference is driven by the lower gas mass in clusters due to
star formation. Our model without star formation or feedback
demonstrates reasonable agreement with the non-radiative simulation at
intermediate scales ($\ell \approx 4000$). However, at larger scales
it lies systematically above the simulation, and below at smaller
scales. The latter may be due to additional power in the simulation
power spectrum due to the presence of substructures. The difference at
large scales is most likely due to an artificial suppression in the
abundance of the most massive clusters due to the limited simulation
volume.

We find that our fiducial model matches that of the AGN simulation at
$\ell = 1800$, but, as with the non-radiative simulation, produces
more power at larger scales and less at smaller scales.  We note that
by increasing the amplitude of the mass-concentration relation by 20\%
our model very nearly reproduces the \citet{battaglia10} simulations
for $\ell > 3000$. The overall reduction between the cases with and
without baryon cooling and feedback are similar for the model and
simulations. This indicates that the feedback prescriptions
incorporated in our fully analytic model are able to reproduce the
results of sub-grid models of AGN feedback in (computationally
intensive) hydrodynamical simulations.

Finally, in the lower-right panel we compare our fiducial model with
that predicted by the \citet{arnaud10} pressure profile. Our model
lies below that inferred by the \citet{arnaud10} profile, particularly
at small scales. As shown in Figure~\ref{fig:Peprofile}, group-mass
objects have a significantly lower pressure at small radii ($r <
\r500$) in our model compared to the A10 profile, which results in
less power at small angular scales. By reducing the level of
non-thermal pressure support in our model by one third (i.e, $\alpha_0
= 0.12$) and setting the redshift evolution of this parameter to zero
our model will produce a power spectrum very similar to that of the
\citet{arnaud10} profile.

\subsection{Comparisons with Observations}

The current generation of SZ surveys (SPT, ACT and Planck) have the
necessary combination of sensitivity, angular resolution, mapping
speed and frequency coverage to make precise measurements of the tSZ
power spectrum.  Multifrequency observations are important for
measuring the tSZ amplitude as both radio point sources and dusty
star-forming galaxies (DSFGs) provide significant foreground signals
near 150 GHz. While the Poisson component of these point source
populations has a different angular dependence to the SZ signal and
can thus be separated and removed, the distribution of DSFGs on the
sky are expected to be clustered with a power spectrum of similar
shape to the SZ signal \citep{viero09,hall10}. Hence, for single
frequency surveys, the clustered DSFG and SZ power spectra cannot be
distinguished. Fortunately, the spectral dependence of these signals
differs greatly and thus observing at two frequencies allows the
clustered DSFG and SZ components to be separated. With three
frequencies the kinetic and tSZ signals can also be independently
measured.

\begin{figure}
\includegraphics[scale = 0.26]{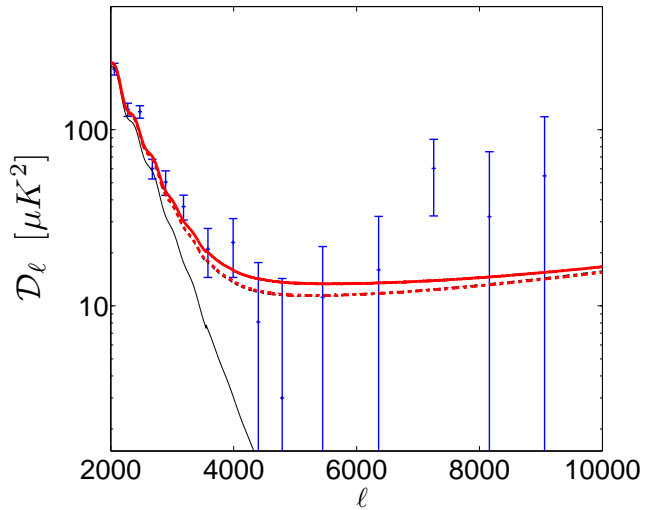}
\caption{Comparisons with SPT measurements of the CMB power spectrum
  at small angular scales. The blue points with error bars are the
  DSFG-removed power spectrum measured by SPT,
  multiplied by 2.2 to preserve the amplitude of the primary CMB power
  spectrum (see text). The dashed black line is the best-fit
  CMB+tSZ+kSZ model to the SPT results.  The red solid line is our
  fiducial tSZ model (plus primary CMB, kSZ and residual point
  sources) with $\sigma_8 = 0.8$, the red dashed line is our fiducial
  model with $\sigma_8 = 0.775$ (almost directly on top of the
  black-dashed line). Results are at 153 GHz.  The thin black line is
  WMAP5 best-fit primary CMB temperature anisotropy power spectrum.}
\label{fig:powerspectra_spt}
\end{figure}

In Figure \ref{fig:powerspectra_spt}, we compare our fiducial model of
the tSZ power spectrum to the recent SPT measurements of the
small-scale CMB power spectrum. In this figure, the blue points with
error bars represent the DSFG-subtracted power spectrum measured by
\citet{lueker10}. This was obtained by combining the maps at 220 and
150 GHz to optimally remove the point source signal. The resulting
DSFG-subtracted power spectrum has been rescaled (by a factor of 2.2)
to preserve the amplitude of the primary CMB signal \citep[see Section
  6.2 of ][]{lueker10}. Therefore, in order to enable a comparison
with the observations we have multiplied the tSZ power spectrum
calculated with our model by this same factor. The dashed black line
is the best-fit CMB+tSZ+kinetic SZ (kSZ) model (plus a residual point
source component) to the SPT observations assuming no theoretical
uncertainty in the SZ predictions.  \citet{lueker10} adopted the
simulations of \citet{sehgal10} to infer a value of $\sigma_8 = 0.746$
from this fit (i.e., from the amplitude of the tSZ power spectrum
alone). The solid and dashed red lines are the tSZ power spectrum
predictions for our fiducial gas model parameters for $\sigma_8 = 0.8$
and $\sigma_8 = 0.775$, respectively (keeping the primary CMB, kSZ and
point source components fixed). As in \citet{lueker10}, we use the
`homogeneous' kinetic SZ power spectrum of \citet{sehgal10}.

Figure \ref{fig:powerspectra_spt} demonstrates that our fiducial model
matches the best-fit SPT power spectrum (dashed black line) when we
lower $\sigma_8$ to 0.775, reducing the discrapency with the WMAP
inferred value and constraints from cluster abundances. Increasing the
values of $\alpha_0$, $\epsfb$ or $\epsdm$ would increase the value of
$\sigma_8$ derived from the SPT measurements further. In general, we
find that our fiducial model scales as $C_\ell \propto \sigma_8^{8.4}$
(at $\ell = 3000$), which is steeper than the power of seven scaling
found by \citet{komatsu02}. This is due to the larger fractional
contribution of higher mass objects to the tSZ power at these scales
in our model, the abundance of which is extremely sensitive to the
value of $\sigma_8$.

\section{Discussion and Conclusion}

The primary goal of this paper is to investigate the level of
theoretical uncertainty on the amplitude of the tSZ angular power
spectrum due to uncertainties in the physics of the ICM. To this end,
we have developed an analytic model for the pressure and density
distribution of the ICM. Our model is based on that of
\citet{ostriker05} which assumes the ICM resides in hydrostatic
equilibrium in the potential well of NFW halos with a polytropic
equation of state. The model accounts for star formation, energy
feedback and energy transfer from dark to gas during mergers.  We have
made a significant improvement to the model by accounting for
non-thermal pressure support in the ICM due to random gas motions,
calibrating the non-thermal pressure profile from hydrodynamical
simulations. This method allows us to rapidly generate theoretical
thermal SZ power spectra, enabling a thorough investigation of the
relative impact of astrophysical and cosmological parameters and their
degeneracies.

Our model has four key free parameters: an energy feedback parameter
$\epsfb$, which determines the amount of non-gravitational heating of
the ICM from supernovae or AGNs, a dark-matter energy transfer
parameter $\epsdm$, which governs the energy exchanged from dark
matter to gas during mergers, a non-thermal pressure support
parameter, $\alpha_0$, which gives the level of non-thermal pressure
at $z = 0$ and its redshift evolution. These parameters were
calibrated against low redshift X-ray observations of groups and
clusters, including the $\m500 - \fgas$ (gas fraction) relation
\citep{vikhlinin06, sun09} and observed electron pressure profiles
\citep{arnaud10}. These studies derive mass estimates for
observed clusters using the equation of hydrostatic equilibrium, which
simulations suggest underestimate true cluster mass by
$10\%-20\%$. Therefore, to ensure a consistent comparison, we are
careful to use the hydrostatic (rather than true) mass predicted by
our model when comparing with observations.

We have found that increasing the energy feedback parameter, $\epsfb$,
from $10^{-7}$ to $10^{-5}$ has little effect on the pressure and
density of high-mass clusters (where feedback energy is a small
fraction of the total binding energy), but produces a factor of 2
decrease in the gas fraction for group mass objects.  Similarly,
increasing the feedback energy dramatically reduces the gas pressure
within $\r500$ and increases the pressure at larger radii in groups.
Raising the level of non-thermal pressure (i.e., increasing
$\alpha_0$) steepens the pressure profiles significantly in the outer
regions ($r \geq \r500$) of both groups and clusters. We found that a
fiducial model of $\epsfb = 10^{-6}$ and $\alpha_0 = 0.18$ provides an
excellent fit to the thermal pressure profile observed by A10.

By combining our gas model with the halo mass function, we
investigated the impact of varying the energy feedback, non-thermal
pressure support and its redshift evolution on the tSZ power
spectrum. We also evaluated the effect of perturbing the normalization
of the mass-concentration relation in our model around the mean
relation reported by \citet{duffy08}. We found that increasing
$\epsfb$ (and $\epsdm$) suppresses power on small angular scales, but
does not strongly affect large-scale power. Hence, the peak of the
power spectrum shifts to larger angular scales as we increase the
amount of energy feedback. This is because $\epsfb$ has a more
significant effect on lower mass systems, which primarily contribute
power on small scales. Varying the normalization of the
mass-concentration relation $A_C$ has a somewhat similar effect; the
power on small scales increases with $A_C$. Raising the concentration
of the host halo deepens the central potential and thus steepens the
gas pressure profile so as to maintain hydrostatic equilibrium.

Varying the non-thermal pressure parameter $\alpha_0$ has the opposite
effect to the feedback parameters; increasing $\alpha_0$ strongly
suppresses power on intermediate and large scales ($\ell < 5000$). The
peak of the power spectrum shifts slightly to smaller scales as
$\alpha_0$ is increased. Generally, we found that non-thermal pressure
support, at the level observed in recent hydrodynamical simulations,
has a significant effect on the tSZ power spectrum, reducing the
amplitude by a factor of 2 relative to the thermal pressure-only
case.

We have demonstrated that our model reproduces the tSZ power spectrum
measured from simulated maps constructed by applying the semi-analytic
model of \citet{bode09} to halos identified in an N-body lightcone
simulation \citep{sehgal10}. This demonstrates that deviations in
halo structural properties (such as scatter in the mass-concentration
relation) do not strongly affect the tSZ power spectrum, justifying
our halo model-based approach. However, comparing with hydrodynamical
simulations demonstrates that analytic models may potentially miss
some small scale power due to substructures in simulations. A more
detailed comparison of hydrodynamical simulations with semi-analytic
models is required to isolate the impact of second order cluster
properties on the SZ power spectrum.

Recent SPT measurements of the small-scale CMB temperature anisotropy
power spectrum \citep{lueker10} have demonstrated there to be some
tension between the value of $\sigma_8$ implied by the measured
amplitude of the tSZ power spectrum at $\ell = 3000$ and those
derived from WMAP observations of the primary CMB power spectrum
\citep{dunkley09, komatsu10}, or from cluster abundances
\citep{vikhlinin09, mantz10}. This tension can be alleviated if the
simulations used by \citet{lueker10} overestimate the amplitude of the
SZ power spectrum by approximately a factor of 2. When compared with
the SPT results, we found that our fiducial model infers $\sigma_8 =
0.775$, and thus reduces the discrepancy in the value of $\sigma_8$
between this and other probes. Generally, we find that our fiducial
model scales as $C_\ell \propto \sigma_8^{8.4}$.

The results presented by \citet{lueker10} encompasses roughly 5\% of
the expected SPT final survey area. Over the next few years, SPT, ACT
and Planck should produce precise measurements of the SZ power
spectrum amplitude over a wide range of angular scales $2000 \leq \ell
\leq 10,000$. We have demonstrated that physics in cluster
environments can modify the shape of the power spectrum, as well as the
amplitude. By comparing the ratio of power measured at small,
intermediate and larger scales it should be possible to derive
information regarding the relative influence of feedback processes and
non-thermal pressure support in the ICM (albeit over a wide range of
mass and redshift). Given that a significant fraction of SZ power
derives from high-redshift and low-mass objects, the tSZ power
spectrum can provide an exciting tool for studying the state of gas in
these objects.

From a theoretical perspective, several improvements can be made to
our modeling of the ICM. First, our prescription assumes feedback
from supernovae and AGNs is even distributed into the ICM. Although a
concerted effort to implement and study AGN feedback in hydrodynamical
simulations of cosmological volumes has only recently begun
\citep{sijacki07,dimatteo08,booth09,teyssier10}, these studies will
shed light on the extent to which non-gravitational heating sources
are able to influence the thermal structure of the ICM. While our
model accounts for energy transfer between dark matter and gas during
mergers, the efficiency with which the process occurs has been poorly
studied in simulations. Furthermore, we have not accounted for any
uncertainty in the slope and amplitude of the stellar mass -- total
mass relation used in our model, which is poorly constrained at higher
redshifts. Finally, the evolution of the radial, mass and redshift
dependence of non-thermal pressure support (from any source) needs to
be studied in hydrodynamical simulations in more detail.  Future X-ray
observatories equipped with high-resolution calorimeters (such as
ASTRO-H) will provide important constraints on the non-thermal
pressure support due to internal gas motions via broadening of heavy
ion emission lines.

In this paper, we have not studied the degeneracies in our model
between astrophysical and cosmological parameters in the shape and
amplitude of the tSZ power spectrum, and how these may limit the
precision to which $\sigma_8$, for example, can currently be
constrained in this way. We leave this to a follow-up paper
\citep{bhattacharya10b}. However, our prescription allows the observed
shape of the power spectrum to be used to weaken or break some of
these degeneracies. Alternatively, one can marginalize over the
feedback or non-thermal pressure parameters in our model in order to
account for astrophysical uncertainty in the predicted tSZ power
spectrum. We intend to make the code for our model publicly available
for use in analyzing observations of the small-scale CMB power
spectrum.

\section{ACKNOWLEDGMENTS}
We thank Gus Evrard, Gil Holder, Jerry Ostriker, Klaus Dolag, Douglas
Rudd, Christian Reichardt and Alexey Vikhlinin for useful
discussions. We also thank Nicholas Battaglia for making the results
of his cosmological hydrodynamics simulations available to us. LS
and DN acknowledge the support of Yale University and NSF grant
AST-1009811. SB acknowledges support from the LDRD and IGPP program at
Los Alamos National Laboratory. This work was supported in part by the
facilities and staff of the Yale University Faculty of Arts and
Sciences High Performance Computing Center.


\end{document}